\documentclass[letterpaper,english,reprint]{revtex4}
\usepackage[T1]{fontenc}
\usepackage[latin9]{inputenc}
\usepackage{refstyle}
\usepackage{amsmath}
\usepackage{esint}

\makeatletter

\AtBeginDocument{\providecommand\secref[1]{\ref{sec:#1}}}
\AtBeginDocument{\providecommand\eqref[1]{\ref{eq:#1}}}
\AtBeginDocument{\providecommand\subref[1]{\ref{sub:#1}}}

\RS@ifundefined{subref}
  {\def\RSsubtxt{section~}\newref{sub}{name = \RSsubtxt}}
  {}
\RS@ifundefined{thmref}
  {\def\RSthmtxt{theorem~}\newref{thm}{name = \RSthmtxt}}
  {}
\RS@ifundefined{lemref}
  {\def\RSlemtxt{lemma~}\newref{lem}{name = \RSlemtxt}}
  {}

\@ifundefined{textcolor}{}
{%
 \definecolor{BLACK}{gray}{0}
 \definecolor{WHITE}{gray}{1}
 \definecolor{RED}{rgb}{1,0,0}
 \definecolor{GREEN}{rgb}{0,1,0}
 \definecolor{BLUE}{rgb}{0,0,1}
 \definecolor{CYAN}{cmyk}{1,0,0,0}
 \definecolor{MAGENTA}{cmyk}{0,1,0,0}
 \definecolor{YELLOW}{cmyk}{0,0,1,0}
}

\makeatother

\usepackage{babel}
\begin{document}

\title{Coordinate-choice independent expression for drift orbit flux and
flux-force relation in neoclassical toroidal viscosity theory }

\author{Q. Peng and M.E. Mauel}

\affiliation{Department of Applied Physics and Applied Mathmatics Columbia University,
New York.}

\address{New York 10027, USA}
\begin{abstract}
A coordinate-choice independent expression does not depend how the
magnetic surface is parametrized by $(\theta,\zeta)$. Flux-force
relation in neoclassical toroidal viscosity(NTV) theory has been generalized
in a coordinate-choice independent way. The expression for the surface
averaged drift orbit flux in $1/\nu$ regime is derived without the
requirement of straight field line coordinates. The resulted formula
is insensitive to how the magnetic surface is parametrized and broadens
the cases where flux-force relation can be applied. Construction of
straight field line coordinates is avoided when the formula is used
for numerical computation.
\end{abstract}
\maketitle

\section{Introduction}

Plasma rotation is of great importance in plasma physics \cite{0029-5515-21-9-003,0029-5515-47-6-S02,1742-6596-123-1-012017,strait:056101,Pustovitov2011}.
Neoclassical toroidal viscosity(NTV) theory is one of the candidates
for explaining the effect on plasma rotation when a non-axisymmetric
perturbation is applied to tokamak \cite{Pustovitov2011,Boozer2009,Shaing2010,Shaing2003,Callen2011,Cole2008}.
Such effect from NTV theory has been observed in many experiments
\cite{PhysRevLett.96.225002,Reimerdes2009,Hua2010}. A fundamental
relation in NTV theory is the flux-force relation, which relates the
neoclassical viscosity forces on plasma to the drift orbit flux through
each magnetic surface \cite{Shaing1996,Shaing2010,0029-5515-21-9-003,RevModPhys.48.239}.
By solving the drift kinetic equation, the drift orbit flux can be
further expressed with respect to the variation of $B$ field on the
magnetic surface \cite{Shaing2003,Shaing2010a}. The drift orbit flux
calculation gives a non-zero result once the axisymmetry is broken
by a 3D perturbation. This whole chain of derivation shows a mechanism
that how the variation of $B$ introduced by perturbation will result
in a surface averaged force that modifies the plasma rotation profile.

Despite the wide usage of such NTV theory, the existing formula for
flux-force relation and the drift orbit flux is not coordinate-choice
independent. By coordinate-choice independent \cite{Isaev1994}, the
authors of this paper mean that the formula does not depend on how
the magnetic surface is parametrized by $(\theta,\zeta)$ --- the
flux label $\psi$ will still be used because the flux surfaces are
determined once the $\vec{B}$ field is given, but the $(\theta,\zeta)$
can be chosen even as non-straight field line coordinates. The flux-force
relation was first derived under Hamada \cite{Shaing1996}, and then
is generalized to other straight field line coordinates (magnetic
coordinates) \cite{Shaing2010}. The expression for drift orbit flux
is only derived under Hamada\cite{Shaing2003}. The flux-force relation
gives the change rate of the surface averaged quantity $\bigl\langle \vec{U}\cdot\vec{Q}\bigr\rangle $,
where $\vec{U}$ is the macro velocity and $\vec{Q}$ a vector field
with certain constrain such that the relation to transport quantity
holds. In the case derived in Hamada \cite{Shaing1996}, it is shown
that $\vec{Q}$ can be the vector field $\vec{e}_{\theta}$, $\vec{e}_{\zeta}$
or a linear combination of them, where $\theta$ and $\zeta$ are
the Hamada coordinates. In the cases derived for other straight field
line coordinates in \cite{Shaing2010}, it has been shown that by
including the pressure term $p$ together with the viscous tensor
$\overleftrightarrow{\pi}$, one can have $\vec{Q}=m\sqrt{g}\vec{\nabla}\psi\times\vec{\nabla}\theta-n\sqrt{g}\vec{\nabla}\psi\times\vec{\nabla}\zeta=m\vec{e}_{\zeta}+n\vec{e}_{\theta}$
and the flux-force relation still holds. Specifically, if $\alpha=m\theta-n\zeta$
is the helical angle for symmetry, $\vec{Q}$ becomes the symmetry
vector. In all these cases shown, the expression for $\vec{Q}$ depends
on how the magnetic surface is parametrized by $(\theta,\zeta)$ and
the coordinates has to be a straight field line coordinate. 

The set of vector fields from which $\vec{Q}$ can be chosen constrains
how the theory can be applied. It is often regarded as a formula for
the NTV toroidal torque if one chooses $\vec{Q}=\vec{e}_{\zeta}$
and $\bigl\langle \vec{U}\cdot\vec{e}_{\zeta}\bigr\rangle $ is treated
as the toroidal angular momentum. Since $\zeta$ has to be one of
the parameters in straight field line coordinates, this equivalence
is not exact for non-axisymmetric cases where $\vec{e}_{\zeta}$ wobbles
and does not agree with the toroidal direction in the lab's frame.
The quasi-symmetry vector $\vec{Q}_{a}$ defined in \cite{Isaev1994}
has another definition $\vec{Q}_{a}=(F\vec{B}+\vec{B}\times\vec{\nabla}\psi)/2\pi B^{2}$,
where $F(\psi)$ is the poloidal electric current flux. This definition
does not depend on the choice of $(\theta,\zeta)$, but the function
$F(\psi)$ is not a free parameter and $\vec{Q}$ thus only represents
one vector field. In this paper, the authors give a coordinate-choice
independent expression for a family of vector fields: $\vec{S}=h(\vec{r})\vec{B}+C(\psi)\vec{B}\times\vec{\nabla}\psi/B^{2}$
which includes all the cases of $\vec{Q}$ shown previously and more.
For any vector fields described by such format, the flux-force relation
will hold. Such format does not depend on how the magnetic surface
is parametrized by $(\theta,\zeta)$ and can be constructed even in
coordinates using lab's toroidal and poloidal angle. 

The expression for drift orbit flux, which is related to $\overleftrightarrow{P}$
term in flux-force relation, will be non-zero when the toroidal symmetry
is broken. This drift orbit flux can be expressed by the variation
of $B$ field on the magnetic surface. The analytic formula has so
far been derived in Hamada \cite{Shaing2010a,Shaing2003}. Using the
vector field $\vec{S}$, the authors generalize the expression of
drift orbit flux for the $1/\nu$ regime case and obtain a coordinate-choice
independent formula that can be evaluated without straight field line
coordinates. 

The results derived in this paper also serve as a potential speed
up method for the numeric calculations since the computation work
for straight field line coordinates construction is avoided. In the
formula, one still needs the flux label $\psi$ for the magnetic field.
The reconstruction codes usually provide such information \cite{Zanca1999,Hanson2009}.

This paper is organized as follows. In \secref{Motivation-of-the},
we briefly introduce how we find the format for $\vec{S}$. In \secref{Proof-of-the},
we prove the coordinate-choice independent flux-force relation using
$\vec{S}$ where the derivation avoids using $\theta$ or $\zeta$
at all. In \secref{Evaluation-of-orbit}, the coordinate-choice independent
expression for the drift orbit flux is derived. In \secref{Comparison-with-existing},
the results are compared with existing theory and some new application
using the generalized results is discussed. Conclusion is made in
\secref{Summary}. In appendix, we discuss the usage of Fourier analysis
on the formula to further speed up the computation.

\section{\label{sec:Motivation-of-the}Motivation of the generalization}

In this section, we take a brief review of the big idea behind flux-force
relation and explain the motivation of finding $\vec{S}$.

The bulk plasma satisfies the momentum equation 
\[
\rho\frac{d\vec{U}}{dt}=\vec{j}\times\vec{B}-\vec{\nabla}\cdot\overleftrightarrow{P}
\]

One can dot this equation with any vector field $\vec{A}$ and obtain
\begin{equation}
\rho\vec{A}\cdot\frac{d\vec{U}}{dt}=\vec{A}\cdot(\vec{j}\times\vec{B})-\vec{A}\cdot(\vec{\nabla}\cdot\overleftrightarrow{P})\label{eq:Momentum_A_general}
\end{equation}

While $\vec{A}$ is arbitrary so far, only certain choices of $\vec{A}$
would make the terms on the R.H.S. of \eqref{Momentum_A_general}
meaningful in the sense that they could be related to transport flux
quantities. It has been shown that in straight field line coordinates
$(\psi,\theta,\zeta)$ \cite{Shaing2010}, if one chooses $\vec{A}=\vec{Q}=m\sqrt{g}\vec{\nabla}\psi\times\vec{\nabla}\theta-n\sqrt{g}\vec{\nabla}\psi\times\vec{\nabla}\zeta$.
Then $\vec{Q}\cdot(\vec{j}\times\vec{B})\propto\vec{j}\cdot\vec{\nabla}\psi$
and $\bigl\langle \vec{A}\cdot(\vec{\nabla}\cdot\overleftrightarrow{P})\bigr\rangle \propto\bigl\langle U_{d}\cdot\vec{\nabla}\psi\bigr\rangle $
where $\vec{U}_{d}$ is the drift orbit velocity and $\bigl\langle X\bigr\rangle $
denotes the surface average of $X$. 

\begin{eqnarray}
\bigl\langle X\bigr\rangle  & = & \oiint X\sqrt{g}d\theta d\zeta\label{eq:surface_avg}\\
 & = & \frac{d}{d\psi}\iiint_{V}Xd^{3}x\nonumber 
\end{eqnarray}

Here, it is not normalized by $V'(\psi)$, because each term in our
final results will be a surface average of some quantity and keeping
$V'(\psi)$ all the way along the derivation only makes it lengthy.
In all, one has \cite{Shaing2010}: 
\begin{eqnarray*}
\rho\frac{d}{dt}\bigl\langle \vec{Q}\cdot\vec{U}\bigr\rangle  & = & (m-nq)\bigl\langle \vec{j}\cdot\vec{\nabla}\psi\bigr\rangle -\bigl\langle \vec{Q}\cdot(\vec{\nabla}\cdot\overleftrightarrow{P})\bigr\rangle \\
\bigl\langle \vec{Q}\cdot(\vec{\nabla}\cdot\overleftrightarrow{P})\bigr\rangle  & = & (m-nq)e\bigl\langle \vec{\Gamma}\cdot\vec{\nabla}\psi\bigr\rangle _{na}\\
 & = & (m-nq)\bigl\langle e\vec{U}_{d}\cdot\vec{\nabla}\psi\bigr\rangle 
\end{eqnarray*}

While all the other quantities are insensitive to how the surface
is parametrized by $(\theta,\zeta)$, the vector field $\vec{Q}=m\vec{e}_{\zeta}+n\vec{e}_{\theta}$
restricts the application to straight field line coordinates and the
vector field $\vec{Q}$ also has to be straight under such coordinates. 

The authors in this paper have relaxed the constrain for $\vec{A}$
in \eqref{Momentum_A_general}. As soon as the flux surfaces are defined
with flux label $\psi$, one can define the perpendicular vector field:
\begin{eqnarray*}
\vec{S}_{0} & = & \frac{\vec{\nabla}\psi\times\vec{B}}{B^{2}}
\end{eqnarray*}

The authors then show in \secref{Proof-of-the} that for any vector
field $\vec{A}=\vec{S}$ which can be casted in the form: 
\begin{equation}
\vec{S}=h(\vec{x})\vec{B}+a(\psi)\vec{S}_{0}\label{eq:general_S_exp}
\end{equation}

where $h(\vec{x})$ is an arbitrary spatial function and $a(\psi)$
is a flux surface function, the terms on the R.H.S. of \eqref{Momentum_A_general}
can be related to transport quantities. As is noted in \cite{Isaev1994},
any vector field tangential to the magnetic surface $\psi(r)=const$
can be represented by:

\[
Q^{*}=C_{0}(\vec{r})\bigl[\frac{\vec{B}\times\vec{\nabla}\psi}{B^{2}}+\lambda(\vec{r})\frac{\vec{B}}{B^{2}}\bigr]
\]
where $C_{0}(\vec{r})$ and $\lambda(\vec{r})$ are free to choose.
By specifying $C_{0}(\vec{r})=1/2\pi$ and letting $\lambda(\vec{r})=F(\psi)$,
the poloidal current flux, one obtains the quasi-symmetry vector $\vec{Q}_{a}$.
The expression for $\vec{S}$ is obtained by only specifying $C_{0}(\vec{r})=C_{0}(\psi)$
while leaving the function $\lambda$ undetermined. Thus, $\vec{Q}_{a}$
is included in the set of $\vec{S}$. In fact, \eqref{general_S_exp}
is equivalent as the condition $\vec{B}\times\vec{S}=a(\psi)\vec{\nabla}\psi$.
For $\vec{Q}=m\vec{e}_{\zeta}+n\vec{e}_{\theta}$ in straight field
line cases where $\vec{B}=q\vec{\nabla}\psi\times\vec{\nabla}\theta+\vec{\nabla}\zeta\times\vec{\nabla}\psi$,
such condition is satisfied with $a(\psi)=m-nq$. Thus, the cases
of $\vec{Q}$ described in \cite{Shaing2010,Shaing1996} are included
in the set of $\vec{S}$.

\section{Proof of the general flux-force relation\label{sec:Proof-of-the}}

In this section, the authors prove the coordinate-choice independent
flux-force relation with the vector field $\vec{A}$ taking the form
in \eqref{general_S_exp}. We first show a few important identities
related to $\vec{S}$ and the guiding center drift velocity. Then
we proceed to prove the relation.

\subsection{An identity for vector field $\vec{S}$.}

Since $\vec{\nabla}\times(a(\psi)\vec{\nabla}\psi)=0$, one has 
\begin{eqnarray*}
 & 0 & =\vec{\nabla}\times(\vec{B}\times\vec{S})\\
 & = & (\vec{\nabla}\cdot\vec{S})\vec{B}+\vec{S}\cdot(\vec{\nabla}\vec{B})-\vec{B}\cdot(\vec{\nabla}\vec{S})
\end{eqnarray*}

Dotting it with $\vec{B}$ one has 
\begin{eqnarray*}
 &  & (\vec{\nabla}\cdot\vec{S})B^{2}-\vec{B}\vec{B}:(\vec{\nabla}\vec{S})\\
 & = & -\vec{S}\cdot\vec{\nabla}(\frac{B^{2}}{2})=-B\vec{S}\cdot\vec{\nabla}B
\end{eqnarray*}

Since $\overleftrightarrow{I}:\vec{\nabla}\vec{S}=\vec{\nabla}\cdot\vec{S}$,
above is equivalent as saying: 
\begin{equation}
(\overleftrightarrow{I}-\vec{b}\vec{b}):(\vec{\nabla}\vec{S})=-\frac{\vec{S}\cdot\vec{\nabla}B}{B}\label{eq:GradS_contraction}
\end{equation}

where $\vec{b}=\vec{B}/B$. This identity will be used many times
in the following derivation.

\subsection{Guiding center drift}

The drift orbit flux $\bigl\langle \vec{U}_{d}\cdot\vec{\nabla}\psi\bigr\rangle $
is the transport quantity we want to relate to. The guiding center
drift velocity of a single particle is\cite{Haseltin}: 
\[
\vec{v}_{D}=\frac{E\times B}{B^{2}}+\frac{m}{qB^{2}}\vec{B}\times\bigl(\frac{\mu}{m}\vec{\nabla}B+v_{\parallel}^{2}\vec{b}\cdot\vec{\nabla}\vec{b}+v_{\parallel}\frac{\partial\vec{b}}{\partial t}\bigr)
\]

where $q$ is the charge of the particle. In the following derivations
in this paper, we will rarely use the safety factor, so $q$ will
mostly be used to denote the particle's charge unless specified otherwise.
Ignoring the polarization drift $\frac{\partial\vec{b}}{\partial t}$
part with semi-steady assumption and noticing that $\vec{E}=-\vec{\nabla}\Phi(\psi)\propto\vec{\nabla}\psi$,
we have

\begin{eqnarray*}
\vec{\nabla}\psi\cdot\vec{v}_{D} & = & \frac{m}{q}\frac{\vec{\nabla}\psi\times\vec{B}}{B^{2}}\cdot\bigl(\frac{\mu}{m}\vec{\nabla}B+v_{\parallel}^{2}\vec{b}\cdot\vec{\nabla}\vec{b}\bigr)\\
 & = & \frac{m}{q}\bigl(\frac{v_{\perp}^{2}}{2B}\vec{S}_{0}\cdot\vec{\nabla}B+v_{\parallel}^{2}\vec{b}\cdot(\vec{\nabla}\vec{b})\cdot\vec{S}_{0}\bigr)
\end{eqnarray*}

Since $\vec{S}_{0}\cdot\vec{b}=0$,

\begin{equation}
\frac{q}{m}\vec{\nabla}\psi\cdot\vec{v}_{D}=\frac{v_{\perp}^{2}}{2B}\vec{S}_{0}\cdot\vec{\nabla}B-v_{\parallel}^{2}\vec{b}\cdot(\vec{\nabla}\vec{S}_{0})\cdot\vec{b}\label{eq:Drift_velocity_1stform}
\end{equation}

For any $\vec{S}$ given by \eqref{general_S_exp} (including $\vec{S}_{0}$),
using \eqref{GradS_contraction} one has 
\begin{eqnarray}
 &  & \frac{v_{\perp}^{2}}{2B}\vec{S}\cdot\vec{\nabla}B-v_{\parallel}^{2}\vec{b}\cdot(\vec{\nabla}\vec{S})\cdot\vec{b}\nonumber \\
 & = & -\frac{v_{\parallel}^{2}-\frac{1}{2}v_{\perp}^{2}}{B}\vec{S}\cdot\vec{\nabla}B-v_{\parallel}^{2}\vec{\nabla}\cdot\vec{S}\nonumber \\
 & = & -\frac{|v_{\parallel}|}{B}\vec{S}\cdot\vec{\nabla}(B|v_{\parallel}(\mu,E,B)|)-v_{\parallel}^{2}\vec{\nabla}\cdot\vec{S}\nonumber \\
 & = & -\frac{|v_{\parallel}|}{B}\vec{\nabla}\cdot(\vec{S}B|v_{\parallel}(\mu,E,B)|)\label{eq:velocity_divergence_form}
\end{eqnarray}

To use above expression, $v_{\parallel}$ must be considered as a
function $v_{\parallel}(\mu,E,B)$ when taking the derivative.

\subsection{R.H.S. terms of \eqref{Momentum_A_general}}

Now, we show the R.H.S. terms of \eqref{Momentum_A_general} can be
related to transport flux quantities. The first term is easy to deal
with: 
\begin{eqnarray*}
\vec{S}\cdot(\vec{j}\times\vec{B}) & = & \vec{j}\cdot(\vec{B}\times\vec{S})=a(\psi)\vec{j}\cdot\vec{\nabla}\psi
\end{eqnarray*}

For the second term, 
\begin{eqnarray*}
\vec{S}\cdot(\vec{\nabla}\cdot\overleftrightarrow{P}) & = & \vec{\nabla}\cdot(\vec{S}\cdot\overleftrightarrow{P})-(\vec{\nabla}\vec{S}):\overleftrightarrow{P}
\end{eqnarray*}
 we notice that 
\begin{eqnarray*}
\overleftrightarrow{P} & = & \sum_{s}\iiint m_{s}f_{s}\vec{v}_{s}\vec{v}_{s}d^{3}v\\
 & = & \sum_{s}\bigl[(\overleftrightarrow{I}-\vec{b}\vec{b})P_{s,\perp}+\vec{b}\vec{b}P_{s,\parallel}\bigr]
\end{eqnarray*}
 where the summation is over different species. The two terms for
$\vec{S}$ in \eqref{general_S_exp} are considered individually.
Using \eqref{GradS_contraction} and \eqref{Drift_velocity_1stform}
we have: 
\begin{eqnarray*}
 &  & (\vec{\nabla}\vec{S}_{0}):\overleftrightarrow{P}{}_{s}=-\frac{\vec{S}_{0}\cdot\vec{\nabla}B}{B}P_{s,\perp}+\vec{b}\vec{b}:(\vec{\nabla}\vec{S}_{0})P_{s,\parallel}\\
 &  & =-\iiint m_{s}\bigl[\frac{1}{2}v_{s,\perp}^{2}\frac{\vec{S}_{0}\cdot\vec{\nabla}B}{B}-v_{s,\parallel}^{2}\vec{b}\vec{b}:(\vec{\nabla}\vec{S}_{0})\bigr]f_{s}d^{3}v\\
 &  & =-q_{s}\iiint v_{d}\cdot\vec{\nabla}\psi f_{s}d^{3}v\\
 &  & =-q_{s}\vec{U}_{s,d}\cdot\vec{\nabla}\psi
\end{eqnarray*}

where $\vec{U}_{s,d}$ is the macro drift orbit velocity for species
$s$. Similarly, using \eqref{GradS_contraction} and \eqref{velocity_divergence_form}
we have:

\begin{widetext}

\begin{eqnarray*}
(\vec{\nabla}h\vec{B}):\overleftrightarrow{P}{}_{s} & = & \iiint\bigl[-\frac{v_{s,\perp}^{2}}{2}\frac{h\vec{B}\cdot\vec{\nabla}B}{B}+v_{s,\parallel}^{2}\vec{b}\cdot(\vec{\nabla}h\vec{B})\cdot\vec{b})\bigr]m_{s}f_{s}d^{3}v\\
 & = & \iiint\frac{|v_{s,\parallel}|}{B}\vec{\nabla}\cdot(h\vec{B}B|v_{s,\parallel}|(\mu,E,B))m_{s}f_{s}d^{3}v\\
 & = & \frac{2\pi}{m_{s}}\iint\bigl\{ \vec{\nabla}\cdot\bigl[h\vec{B}B|v_{s,\parallel}|(\mu,E,B)f_{s}\bigr]-|v_{s,\parallel}|hB\vec{B}\cdot\nabla f_{s}\bigr\} d\mu dE\\
 & = & \vec{\nabla}\cdot(P_{s,\parallel}h\vec{B})-\frac{2\pi}{m_{s}}\iint|v_{s\parallel}|hB\vec{B}\cdot\vec{\nabla}f_{s}(\mu,E)d\mu dE
\end{eqnarray*}

In all, since $\vec{S}=h\vec{B}+a\vec{S}_{0}$, we have:

\[
\rho\vec{S}\cdot\frac{d\vec{U}}{dt}=a\vec{j}\cdot\nabla\psi-\vec{\nabla}\cdot(\vec{S}\cdot\overleftrightarrow{P})-a\sum_{s}q_{s}\vec{U}_{s,d}\cdot\vec{\nabla}\psi+\vec{\nabla}\cdot(P_{\parallel}h\vec{B})-\sum_{s}\frac{2\pi}{m_{s}}\iint|v_{s,\parallel}|hB\vec{B}\cdot\vec{\nabla}f_{s}(\mu,E)d\mu dE
\]

\end{widetext}

Since $\bigl\langle \vec{\nabla}\cdot\vec{X}\bigr\rangle =\frac{d}{d\psi}\bigl\langle \vec{X}\cdot\vec{\nabla}\psi\bigr\rangle $
and $\vec{b}\cdot\vec{\nabla}\psi=0$, $\vec{S}\cdot\vec{\nabla}\psi=0$,
those total divergence terms should vanish after surface average.
If we also take into account that under low collisionality regime,
$\vec{b}\cdot\vec{\nabla}f_{s}(\mu,E)=0$ \cite{Shaing1996,Shaing1983},
then we obtain: 
\begin{equation}
\bigl\langle \rho\vec{S}\cdot\frac{d\vec{U}}{dt}\bigr\rangle =a\bigl\langle \vec{j}\cdot\nabla\psi\bigr\rangle -a\sum_{s}\bigl\langle q_{s}\vec{U}_{s,D}\cdot\vec{\nabla}\psi\bigr\rangle \label{eq:flux_force_relation}
\end{equation}

Such flux-force relation, where $\vec{S}$ is expressed by \eqref{general_S_exp},
is insensitive to how the magnetic surface is parametrized by $(\theta,\zeta)$.

\section{Evaluation of the drift orbit flux\label{sec:Evaluation-of-orbit}}

The formula for evaluating the drift orbit flux $\bigl\langle q_{s}\vec{U}_{s,D}\cdot\vec{\nabla}\psi\bigr\rangle $
has been derived under Hamada coordinates \cite{Shaing2003} and this
quantity should be a surface invariant that does not depend on the
choice of $(\theta,\zeta)$ \cite{Shaing2010}. So technically, one
already has a method to calculate $\bigl\langle \rho\vec{S}\cdot\frac{d\vec{U}}{dt}\bigr\rangle $.
However, the derivation in the previous section encourages the authors
to find a way of representing the quantity $\bigl\langle q_{s}\vec{U}_{s,D}\cdot\vec{\nabla}\psi\bigr\rangle $
without Hamada. A coordinate-choice independent formula for this transport
flux under $1/\nu$ regime will be derived in this section, which
can be evaluated under much more general coordinates. The drift orbit
flux under other regimes \cite{Shaing2012,Shaing2010a,Sun2011} are
not covered in this paper.

\subsection{Bounce average\label{sub:Bounce-average}}

Bounce integral $\oint dl$ is one of the key elements of the derivations
in this section. Operators like $\partial/\partial\mu$ or $\int d\mu$
will be moved in/out the bounce integral and the surface average will
also be decomposed into $\oint dl$ and the integral over the field
line label. On a perturbed magnetic surface where the trapping region
possibly contains holes in it, these tricks could be confusing and
usage of the formula could raise errors in numerical computation if
the definitions are not clear. The authors try to clarify such issues
in this subsection. The derivation will use the Clebsch coordinate
$(l,\beta,\psi)$, but the $\oint dl$ integral we adapt in the end
is an integral along field lines that does not depend on this coordinate
as long as the $\vec{B}$ is properly defined.

The concept of trapped/passing particles are introduced when exchanging
the order of integral over velocity and space. The velocity space
$d^{3}v$ can also be represented by $d\mu dEd\gamma$ \cite{Haseltin}
and the $\mu$ component only spans over $[0,E/B(\vec{x})]$. Consider
a function $h(\theta,\zeta,E,\mu)$ that depends on both space and
velocity. One can integrate this function over the velocity space
first and then further integrate it over the whole surface. 
\[
I=\sum_{\sigma}\int_{0}^{2\pi}\int_{0}^{2\pi}\int_{0}^{\infty}\int_{0}^{E/B(x)}hd\mu dEd\theta d\zeta
\]

where $\sigma=\pm1$ denotes the direction of $v_{\parallel}$ for
the given $(\mu,E)$\cite{Haseltin}, which will be omitted for most
of the derivation. If one integrates it over $\theta-\zeta$ domain
first, then the maximum $\mu$ it can reach is $E/B_{min}$ and for
a given $(E,\mu)$ the integration area for $\theta-\zeta$ may not
cover the whole surface: 
\[
I=\int_{0}^{\infty}\int_{0}^{E/B_{min}}\int_{0}^{2\pi}\int_{\theta_{1}(\mu,E,\zeta)}^{\theta_{2}(\mu,E,\zeta)}hd\theta d\zeta d\mu dE
\]

$\theta_{1}$ and $\theta_{2}$ will be the turning point where $B(\theta,\zeta)=E/\mu$.
If $E/\mu>B_{max}$ ,then $\theta_{1}=0$ and $\theta_{2}=2\pi$ and
the area is still the whole $\theta-\zeta$ domain. Another way of
saying this is, for a given $(\mu,E)$ it determines a trapping domain
$\Omega(E,\mu)$ on the surface where $\mu B(\vec{x})<E$ is satisfied
within this domain. The domain $\Omega(E,\mu)$ is the area on which
we should integrate $d\theta d\zeta$. 

Bounce integral is along field lines. If we define an indicator function
$1_{\Omega}$, the value of which is one within the domain $\Omega$
and zero outside it, then the integration over $\Omega$ could be
expressed as: 
\begin{eqnarray}
 &  & \int_{0}^{2\pi}\int_{\theta_{1}(\mu,E,\zeta)}^{\theta_{2}(\mu,E,\zeta)}h\sqrt{g}d\theta d\zeta\nonumber \\
 & = & \int_{0}^{2\pi}\int_{0}^{2\pi}1_{\Omega}h\sqrt{g}d\theta d\zeta=\iint1_{\Omega}h\frac{dl}{B}d\beta\label{eq:dldbeta_surface}
\end{eqnarray}

where $\sqrt{g}$ is the Jacobian. If the integration $\int dl$ is
intersected by $1_{\Omega}$, then for each continuous pieces along
the line, it becomes a bounce integration $\oint dl$ on the field
line once $\sum_{\sigma}$ is considered. Sometimes the $dl$ integration
has to go several rounds over the torus before it meets the edge of
$\Omega$. Yet the domain of $\iint dld\beta$ in \eqref{dldbeta_surface}
only covers the surface one time and we may not get a complete $\oint dl$.
Also, while intuition suggests $\oint dl$ for a passing particle
should sample the whole surface, the relation is still not clear.

To clarify this issue and obtain a more uniform expression, we need
to increase the domain of $l$ to cover the surface multiple times.
If $(l,\beta)$ is used to parametrize the surface, then each point
on the surface can actually be represented by multiple $(l,\beta)$
points. Suppose $A_{1}$ is one simple connected domain for $(l,\beta)$
such that each point on the surface is exactly represented once in
$A_{1}$ and the lower boundary for $l$ in $A_{1}$ is uniformly
$0$. Let the span for $\beta$ in $A_{1}$ be $(0,\beta_{0})$. The
upper boundary for $l$, would then be a function of $\beta$: $L_{1}(\beta)$,
which indicates the length of each $B$ field line in $A_{1}$ before
it meets the boundary. If we keep the span of $\beta$ in $A_{1}$
fixed, but increase the upper boundary for $l$, then we can construct
$A_{2}$, which has every point on the surface being mapped exactly
to two points in $A_{2}$. We thus have $L_{2}(\beta)$, which allows
the field line to go around the surface exactly twice. Similarly,
we can define $A_{N}$ and $L_{N}(\beta)\rightarrow\infty$ as $N\rightarrow\infty$.
Then because of the periodicity of the surface, the surface integral
has the following property 
\[
\iint h\sqrt{g}dld\beta=\iint_{A_{1}}h\sqrt{g}dld\beta=\frac{1}{N}\iint_{A_{N}}h\sqrt{g}dld\beta
\]

Since the domain for $\beta$ is $(0,\beta_{0})$, the surface integration
becomes:

\[
\iint h\sqrt{g}dld\beta=\int_{0}^{\beta_{0}}\frac{1}{N}\int_{0}^{L_{N}(\beta)}h\sqrt{g}dld\beta
\]

If one thinks it in this way for the integration in \eqref{dldbeta_surface}.
Then for trapped particles, $\frac{1}{N}\int_{0}^{L_{N}(\beta)}1_{\Omega}hdl$
is effectively cut into several pieces divided by the edge of $\Omega$
and each piece forms a $\frac{1}{N}\oint hdl$ with the $\sum_{\sigma}$
taken into consideration. For some segments containing the end points
$l=0$ or $l=L_{N}(\beta)$, it may not form a complete bounce circle,
but we can always add the missing parts. As $N\rightarrow\infty$,
the difference, scaled by $1/N$, goes to zero and the final expression
converges to the same value whether completing the circle or not on
the boundary. For each bouncing circle on the surface, since the surface
is repeated $N$ times in $A_{N}$, it will occur $N$ times in the
integration once sums over $(0,\beta_{0})$. Thus, the total effective
value for each bouncing integration will be still be $N\times\frac{1}{N}\oint hdl=\oint hdl$. 

For passing particles, $\frac{1}{N}\int1_{\Omega}hdl$ is $\frac{1}{N}\int_{0}^{L_{N}(\beta)}hdl$.
We can still denote this integration by $\oint hdl$. This is especially
important if $h$ has the form of $h=\sigma\partial H/\partial l$
where $H$ is a periodic function defined on the surface. In such
cases, suppose the maximum value of $|H|$ on the surface is $H_{M}$,
then $|\frac{1}{N}\int_{0}^{L_{N}(\beta)}hdl|<\frac{2}{N}H_{M}$ which
goes to $0$ as $N\rightarrow\infty$. Considering the trivial trapped
particle case together, we have that $\oint\sigma[\partial H/\partial l]dl=0$
for both passing and trapped particles.

In all, for a integration over the surface $\iint1_{\Omega}h\frac{dl}{B}d\beta$,
it can be reformed as an integration over $A_{N}$ and take the limit
of $N\rightarrow\infty$. 
\begin{eqnarray}
 &  & \iint1_{\Omega}h\frac{dl}{B}d\beta=\frac{1}{N}\iint_{A_{N}}1_{\Omega}h\frac{dl}{B}d\beta\nonumber \\
 &  & =\int_{0}^{\beta_{0}}\frac{1}{N}\int_{0}^{L_{N}(\beta)}1_{\Omega}h\frac{dl}{B}d\beta\nonumber \\
 &  & =\lim_{N\rightarrow\infty}\int_{0}^{\beta_{0}}\sum_{k}(\frac{1}{N}\oint_{k}h\frac{dl}{B})d\beta\label{eq:surface_bounce_formula}
\end{eqnarray}

where $\oint_{k}dl$ is the $kth$ continuous piece on $[0,L_{N}(\beta)]$
that has been cut by the boundary of $\Omega$. As $N$ increases,
the total number of $k$ for each $\beta$ also increases. The total
weight for each bouncing circle on the surface is still one. We can
omit the summation and $1/N$ notation together if no ambiguity would
arise. And we have $\iint1_{\Omega}h\frac{dl}{B}d\beta=\int(\oint h\frac{dl}{B})d\beta$. 

It is not wise to numerically evaluate the surface average as a limit
shown in \eqref{surface_bounce_formula}. This relation, as will be
seen later, serves as a transient step for the derivation. In the
final expression shown in \eqref{orbit_drift_flux_expr}, the surface
average can be evaluated using \eqref{surface_avg}. In the circumstances
shown in \subref{Fast-calculation-by}, $1_{\Omega}$ is well contained
in $A_{1}$ and one can just use \eqref{surface_bounce_formula} for
the surface average and take $N=1$.

\subsection{Drift orbit flux}

The expression for the drift orbit flux is obtained by solving the
drift kinetic equation. The procedure of obtaining the quadratic $\partial f/\partial\mu$
expression is very similar to the work in \cite{Shaing2003} which
uses Hamada. We present it here in a modified way to show that it
is actually coordinate-choice independent. Then we proceed to solve
for $\partial f/\partial\mu$ and express that with vector field $\vec{S}$.
The final coordinate-choice independent expression provides a method
for numeric calculation that does not need straight field line coordinates.

The 0th order drift kinetic equation gives a Maxwellian solution $f_{M}$
for $f_{0}$. The first order equation is \cite{Shaing2003}: 
\[
v_{\parallel}\vec{b}\cdot\vec{\nabla}f_{1}+\vec{v}_{d}\cdot\vec{\nabla}(f_{0}+f_{1})=C[f_{1}]
\]

Further expansion of this equation with $1/\nu$ ordering suggests:

\begin{equation}
v_{\parallel}\vec{b}\cdot\vec{\nabla}f_{1,0}=0\label{eq:f_10_equation}
\end{equation}

\begin{equation}
v_{\parallel}\vec{b}\cdot\vec{\nabla}f_{1,1}+\vec{v}_{d}\cdot\vec{\nabla}\psi\frac{df_{0}}{d\psi}=\frac{v_{\parallel}}{B}\frac{\partial}{\partial\mu}\bigl(\nu M\mu v_{\parallel}\frac{\partial f_{1,0}}{\partial\mu}\bigr)\label{eq:partialf_10_overmu}
\end{equation}

The first equation implies $f_{1,0}=f_{1,0}(E,\mu,\psi,\beta)$ does
not vary along the field line. Although two different lines are essentially
one field line on irrational surface, $f_{1,0}$ can have different
value on two field lines for trapped particles with the same $(E,\mu)$.
This is because those two field lines are disconnected by the boundaries
of trapping region $\Omega(E,\mu)$. $v_{\parallel}$ is not properly
defined outside the trapping region and \eqref{f_10_equation} is
not valid outside the trapping region. So, the value of $f_{1,0}$
on this two pieces of lines does not talk to each other. 

The second equation is on both the spatial and velocity domain $(\beta,l,E,\mu)$.
For a given $(E,\mu)$, one knows whether it is a passing particle
or trapped particle. If it is a trapped particle, then the spatial
point $(\beta,l)$ corresponds to a specific $\oint dl$ whose integration
domain includes this point. If it is a passing particle, then it corresponds
to the $\oint dl$ with $l$ goes from $0$ to $L_{N}(\beta)$, which
essentially samples the whole surface as $N\rightarrow\infty$ . In
either case, one can multiply \eqref{partialf_10_overmu} with $1/|v_{\parallel}|$
and perform the $\oint dl$ integration. The first term on the L.H.S
takes the form $\oint\sigma[\partial H/\partial l]dl$ and will vanish
for both passing and trapped cases, as is discussed in \subref{Bounce-average}.
Thus we have: 
\begin{eqnarray}
 &  & \bigl(\oint\frac{B}{|v_{\parallel}|}\vec{v}_{d}\cdot\vec{\nabla}\psi\frac{dl}{B}\bigr)\frac{df_{0}}{d\psi}\nonumber \\
 & = & \oint\frac{\partial}{\partial\mu}\bigl(\nu M\mu|v_{\parallel}|\frac{\partial f_{1,0}}{\partial\mu}\bigr)\frac{dl}{B}\nonumber \\
 & = & \frac{\partial}{\partial\mu}\bigl(\oint\nu M\mu|v_{\parallel}|\frac{\partial f_{1,0}}{\partial\mu}\frac{dl}{B}\bigr)\label{eq:partial_mu}
\end{eqnarray}

Moving the operator $\partial/\partial\mu$ outside $\oint dl$ in
the last step of deriving needs more explanation because the domain
for $dl$ depends on $\mu$. Suppose the equation is used for a given
$\mu_{1}$, then moving $\oint dl$ inside requires a proper definition
for the integral domain for at least $\mu\in[\mu_{1},\mu_{1}+\epsilon)$,
where $\epsilon$ is just a small quantity. For example, if the integral
of $\oint dl$ is on $[l_{1},l_{2}]$ for $\mu_{1}$, while for $\mu_{1}+\epsilon$,
$|B|$ has reached some critical point within the region and the domain
$[l_{1},l_{2}]$ has to be broken into several pieces, then the $\oint dl$
for $\mu_{1}+\epsilon$ has to be actually a summation $\sum_{k}\oint dl$
so that all the pieces in $[l_{1},l_{2}]$ are added to make the last
step in \eqref{partial_mu} valid.

The drift orbit flux from $f_{0}$ does not contribute when surface
averaged. The first non-zero contribution comes from the $f_{1,0}$
part. As has been noted in the previous section, the integration over
the surface can also be represented as the limit of integration over
$A_{N}$ and divided by $N$. This makes sure that when $\oint dl$
is performed, it will be a complete circle. By exchanging the order
of integration, we have:

\begin{eqnarray*}
 &  & q\Gamma=\bigl\langle \int d^{3}vq\vec{v}_{d}\cdot\vec{\nabla}\psi f_{1,0}\bigr\rangle \\
 &  & =\bigl\langle \frac{2\pi q}{M^{2}}\iint\frac{B}{|v_{\parallel}|}f_{1,0}\vec{v}_{d}\cdot\vec{\nabla}\psi d\mu dE\bigr\rangle \\
 &  & =\frac{2\pi q}{M^{2}}\iint_{A_{N}}\frac{1}{N}\bigl[\iint\frac{B}{|v_{\parallel}|}\vec{v}_{d}\cdot\vec{\nabla}\psi f_{1,0}d\mu dE\bigr]\frac{dl}{B}d\beta\\
 &  & =\frac{2\pi q}{M^{2}}\iint(\int\oint\frac{B}{|v_{\parallel}|}\vec{v}_{d}\cdot\vec{\nabla}\psi\frac{dl}{B}f_{1,0}d\beta)d\mu dE
\end{eqnarray*}

Then substitute in \eqref{partial_mu} and exchange the order of integration
again, we obtain the expression with respect to $(\partial f_{1,0}/\partial\mu)^{2}$:
\begin{eqnarray}
 &  & q\Gamma=-\frac{2\pi q}{M^{2}}\iint\int\frac{(\oint\nu M\mu|v_{\parallel}|\frac{dl}{B})\frac{\partial f_{1,0}}{\partial\mu}\frac{\partial f_{1,0}}{\partial\mu}}{df_{0}/d\psi}d\beta d\mu dE\nonumber \\
 &  & =-\frac{2\pi q}{M}\bigl\langle \iint\nu\mu|v_{\parallel}|\bigl(\frac{\partial f_{1,0}}{\partial\mu}\bigr)^{2}\bigl(\frac{df_{0}}{d\psi}\bigr)^{-1}d\mu dE\bigr\rangle \label{eq:flux_quadratic_form}
\end{eqnarray}

where the surface integral has been treated the same as that in deriving
\eqref{surface_bounce_formula} and we have used \eqref{f_10_equation}
to assume that $f_{1,0}$ and $\partial f_{1,0}/\partial\mu$ can
be moved inside/outside the integral $\oint dl$.

\subsection{Solution for $\partial f_{1,0}/\partial\mu$}

Now, we have obtained the quadratic $\partial f_{1,0}/\partial\mu$
expression for the drift orbit flux. It remains to solve for $\partial f_{1,0}/\partial\mu$.
This can be done by substituting \eqref{Drift_velocity_1stform} and
\eqref{velocity_divergence_form} into \eqref{partial_mu}:

\begin{eqnarray}
 &  & q\frac{\partial}{\partial\mu}\bigl(\oint\nu\mu|v_{\parallel}|\frac{dl}{B}\frac{\partial f_{1,0}}{\partial\mu}\bigr)\nonumber \\
 & = & \bigl(\oint-\vec{\nabla}\cdot(\vec{S}_{0}B|v_{\parallel}(\mu,E,B)|)\frac{dl}{B}\bigr)\frac{df_{0}}{d\psi}\label{eq:partial_mu_solve_1}
\end{eqnarray}

Integration by $\mu$,
\begin{eqnarray}
 &  & q\oint\nu\mu|v_{\parallel}|\frac{dl}{B}\frac{\partial f_{1,0}}{\partial\mu}\label{eq:partial_mu_solve_2}\\
 & = & -\int_{\mu_{max}}^{\mu}\bigl(\sum_{k}\oint_{k}\vec{\nabla}\cdot(\vec{S}_{0}B|v_{\parallel}(\tilde{\mu},E,B)|)\frac{dl}{B}\bigr)\frac{df_{0}}{d\psi}d\tilde{\mu}\nonumber 
\end{eqnarray}

The discussion about the last step in \eqref{partial_mu} makes sure
that the $\oint dl$ we obtain on the L.H.S. is consistent with our
previous definitions for $\oint dl$. There's no constant term because
L.H.S also equals $0$ at $\mu=\mu_{\max}$. The integration limit
of $\oint dl$ on the R.H.S. is given by $(E,\tilde{\mu})$. As has
been noted in previous section, when integrating \eqref{partial_mu_solve_1},
one has to connect all the locally defined equation. Suppose \eqref{partial_mu_solve_2}
is for $\mu=\mu_{1}$ and $[l_{1},l_{2}]$ is the corresponding domain
for $\oint dl$. Then for $\tilde{\mu}>\mu_{1}$ on the R.H.S. of
\eqref{partial_mu_solve_2}, it has to include all the pieces of $\oint dl$
within $[l_{1},l_{2}]$. 

Moving the $\oint dl$ outside $\int d\tilde{\mu}$ integration will
affect the domain of $\oint dl$, which will become the largest for
all $\tilde{\mu}\in[\mu,\mu_{max}]$. The integration domain for $d\mu$
will also be affected. At each spacial point $x$, the integration
domain for $d\mu$ will be $[\mu,\tilde{\mu}_{max}]$ where $\tilde{\mu}_{max}=E/B(x)$.

\begin{eqnarray*}
 &  & -\int_{\mu_{max}}^{\mu}\bigl(\oint\vec{\nabla}\cdot(\vec{S}_{0}B|v_{\parallel}(\tilde{\mu},E,B)|)\frac{dl}{B}\bigr)\frac{df_{0}}{d\psi}d\tilde{\mu}\\
 & = & \oint\vec{\nabla}\cdot(\vec{S}_{0}B\int_{\mu}^{\tilde{\mu}_{max}}|v_{\parallel}(\tilde{\mu},E,B)|d\tilde{\mu})\frac{dl}{B}\frac{df_{0}}{d\psi}\\
 & = & \frac{m}{3}\oint\vec{\nabla}\cdot(\vec{S}_{0}|v_{\parallel}|^{3})\frac{dl}{B}\frac{df_{0}}{d\psi}
\end{eqnarray*}

This procedure has no problem of including the passing particle case,
where $\oint Xdl$ is $\frac{1}{N}\int_{0}^{L_{N}(\beta)}Xdl$. Thus
we have: 
\begin{eqnarray*}
 &  & \frac{\partial f_{1,0}}{\partial\mu}=\frac{m\oint\vec{\nabla}\cdot(\vec{S}_{0}|v_{\parallel}|^{3})\frac{dl}{B}}{3q\nu\mu\oint|v_{\parallel}|\frac{dl}{B}}\frac{df_{0}}{d\psi}\\
 &  & =\frac{\oint\bigl(|v_{\parallel}|\vec{S}_{0}\cdot\vec{\nabla}B-\frac{m|v_{\parallel}|^{3}}{3\mu}\vec{\nabla}\cdot\vec{S}_{0}\bigr)\frac{dl}{B}}{q\nu\oint|v_{\parallel}|\frac{dl}{B}}\frac{df_{0}}{d\psi}
\end{eqnarray*}

In the case of $\vec{S}=a\vec{S}_{0}+h\vec{B}$, for trapped particles
$|v_{\parallel}|=0$ at the turning point and 
\[
\oint\vec{\nabla}\cdot(h\vec{B}|v_{\parallel}|^{3})\frac{dl}{B}=\oint\vec{b}\cdot\vec{\nabla}(h|v_{\parallel}|^{3})dl=0
\]

Thus

\begin{equation}
\frac{\partial f_{1,0}}{\partial\mu}=\frac{m\oint\vec{\nabla}\cdot(\vec{S}|v_{\parallel}|^{3})\frac{dl}{B}}{3aq\nu\mu\oint|v_{\parallel}|\frac{dl}{B}}\frac{df_{0}}{d\psi}\label{eq:partial_mu_solution}
\end{equation}

This is also valid for passing particles because $\oint\vec{b}\cdot\vec{\nabla}(h|v_{\parallel}|^{3})dl$
is bounded by $Max(2h|v_{\parallel}|^{3})/N$, which goes to zero
as $N,\ L_{N}(\beta)\rightarrow\infty$, while $\oint|v_{\parallel}|\frac{dl}{B}$
stays finite. Thus the $h\vec{B}$ part still does not contribute
in \eqref{partial_mu_solution}. Finally, we obtain the formula for
the surface integrated drift orbit flux. 

\begin{widetext}

\begin{eqnarray}
 &  & \bigl\langle qU_{d}\cdot\vec{\nabla}\psi\bigr\rangle =-\frac{2\pi m}{9a^{2}q}\bigl\langle \iint\frac{1}{\mu\nu}|v_{\parallel}|\bigl(\frac{\oint\vec{\nabla}\cdot(\vec{S}|v_{\parallel}|^{3})\frac{dl}{B}}{\oint|v_{\parallel}|\frac{dl}{B}}\bigr)^{2}\bigl(\frac{df_{0}}{d\psi}\bigr)d\mu dE\bigr\rangle \label{eq:orbit_drift_flux_expr}\\
 &  & =-\frac{2\pi}{a^{2}qm}\bigl\langle \iint\frac{\mu}{\nu}|v_{\parallel}|\bigl(\frac{\oint\bigl(|v_{\parallel}|\vec{S}\cdot\vec{\nabla}B-\frac{m|v_{\parallel}|^{3}}{3\mu}\vec{\nabla}\cdot\vec{S}\bigr)\frac{dl}{B}}{\oint|v_{\parallel}|\frac{dl}{B}}\bigr)^{2}\bigl(\frac{df_{0}}{d\psi}\bigr)d\mu dE\bigr\rangle \label{eq:orbit_drift_flux_expr_2ndform}
\end{eqnarray}

\end{widetext}

This expression is coordinate-choice independent and can be evaluated
even in non-straight field line coordinates. For trapped particles,
in order to perform $\oint dl$, one needs to calculate the boundaries
of trapping region, which are the contours given by $|B|=E/\mu$.
For passing particles, $\oint\frac{dl}{B}$ effectively becomes the
surface integration. Since $\bigl\langle \vec{\nabla}\cdot(\vec{S}|v_{\parallel}|^{3})\bigr\rangle =0$,
it means that the passing particles do not contribute. 

This expression, combined with the flux-force relation in \eqref{flux_force_relation},
gives a way to calculate the effect of a 3D magnetic field perturbation
on plasma rotation. It should be noted that the $\vec{S}$ in \eqref{orbit_drift_flux_expr}
does not need to be the same $\vec{S}$ as that in \eqref{flux_force_relation}.
One can use $\vec{S}_{1}$ to calculate the drift orbit flux and use
$\vec{S}_{2}$ in \eqref{flux_force_relation} to calculate the change
rate of $\bigl\langle \vec{U}\cdot\vec{S}_{2}\bigr\rangle $, as
long as they all follow the general expression for $\vec{S}$ in \eqref{general_S_exp}.

\section{Comparison with existing theories and applications\label{sec:Comparison-with-existing}}

\subsection{Flux-force relation}

One can choose $\vec{S}$ to be many vector fields of interest, as
long as it can be cast into the form expressed by \eqref{general_S_exp},
which is equivalent as requiring $\vec{B}\times\vec{S}=a(\psi)\vec{\nabla}\psi$.
The vector field $\vec{e}_{\theta}$, $\vec{e}_{\zeta}$ in \cite{Shaing1996},
$\vec{Q}$, $\vec{B}_{t}$ and $\vec{B}_{p}$ defined in \cite{Shaing2010}
all satisfy such requirements. 

For example, in straight field line cases: $\vec{B}=\frac{1}{\sqrt{g}}(\vec{e}_{\zeta}+q\vec{e}_{\theta})$,
where $q$ is the safety factor. If one takes $\vec{S}=m\sqrt{g}\vec{\nabla}\psi\times\vec{\nabla}\theta-n\sqrt{g}\vec{\nabla}\psi\times\vec{\nabla}\zeta$,
then $\vec{B}\times\vec{S}=(m-nq)\vec{\nabla}\psi$. This is sufficient
to say that $\vec{S}=h\vec{B}+(m-nq)\vec{S}_{0}$, where the explicit
expression for $h(\vec{x})$ is yet to be calculated. Using such $\vec{S}$
in \eqref{flux_force_relation}, one recovers the results obtained
in \cite{Shaing1996,Shaing2010}. 

The relation is equally useful in non-straight field line coordinates
$(\theta,\zeta,\psi)$, where $\psi$ is still the flux label and
\[
\vec{B}=\frac{1}{\sqrt{g}}\bigl[B_{t}(\theta,\zeta,\psi)\vec{e}_{\zeta}+B_{p}(\theta,\zeta,\psi)\vec{e}_{\theta}\bigr]
\]
where $B_{t}$ and $B_{p}$ are no longer required to be constant
on the magnetic surface. One can construct a subset of $\vec{S}$
with:

\[
\vec{S}=\frac{a_{1}(\psi)}{B_{p}}\vec{e}_{\zeta}+\frac{a_{2}(\psi)}{B_{t}}\vec{e}_{\theta}
\]
 such that $\vec{B}\times\vec{S}=(a_{1}-a_{2})\vec{\nabla}\psi$ is
always satisfied. Especially, if one further specifies $a_{2}(\psi)=0$,
then $\vec{S}=a_{1}\vec{e}_{\zeta}/B_{p}$. Since there is no requirement
on $(\theta,\zeta)$, one can choose them to be the natural toroidal
and poloidal angles. $\vec{S}$ obtained with such choice is close
to the lab's toroidal direction, but still not exactly the same. This
explicitly shows that in non-axisymmetric cases, $\bigl\langle \rho\vec{U}\cdot\vec{S}\bigr\rangle $
is not really the toroidal angular momentum, but with some modulation
by $B_{p}$ and the wobbliness on magnetic surface.

\subsection{drift orbit flux}

The drift orbit flux formula has previously been derived in Hamada
$(\theta,\zeta,V)$. If one chooses $\vec{S}=\vec{e}_{\zeta}$, then
$\vec{\nabla}\cdot\vec{S}=0$ and the corresponding term in \eqref{orbit_drift_flux_expr_2ndform}
is annihilated. One gets the result consistent with that obtained
in \cite{Shaing2003}.

The surface average is performed on the flux surface after perturbation.
In deriving \eqref{orbit_drift_flux_expr}, we didn't make the assumption
about the normalized perturbation amplitude being much smaller than
$\epsilon$, the inverse aspect ratio. Without this assumption, the
bounce integral on different field lines could not be factored out
since the turning point is modified by the 3D perturbation once $\delta B/B$
is comparable to $\epsilon$. If we further adapt this assumption
as that in \cite{Shaing2003,Shaing2010a}, then \eqref{orbit_drift_flux_expr}
could be further simplified by using Fourier analysis and performing
the integration with respect to the field line label $\beta$ separately
from the bounce integral. This is shown in \subref{Fast-calculation-by}.
The authors want to keep the formula general to include stellerator-like
perturbed cases as well.

The formula given by \eqref{orbit_drift_flux_expr} or \eqref{orbit_drift_flux_expr_2ndform}
can be used under other straight field line coordinates and even non-straight
field line coordinates. One still needs to determine the trapping
region $\Omega(E,\mu)$ for each given $(E,\mu)$, the boundaries
of which are defined by the contours: $|B|=E/\mu$. This is a necessary
work for calculating $\oint dl$ related integrals.

\section{Summary\label{sec:Summary}}

The authors show that the flux-force relation expressed by \eqref{flux_force_relation}
is valid as long as the vector field $\vec{S}$ can be cast into the
coordinate-choice independent form expressed by \eqref{general_S_exp}.
This broadens the application of flux-force relation to a bigger set
of $\vec{S}$ and drops the requirement of straight field line coordinates. 

The surface averaged drift orbit flux $\bigl\langle q_{s}\vec{U}_{s,D}\cdot\vec{\nabla}\psi\bigr\rangle $,
which is directly related to the flux-force formula, is also expressed
in a coordinate-choice independent way with respect to the variation
of $|B|$ under $1/\nu$ regime. The evaluation of \eqref{orbit_drift_flux_expr}
or \eqref{orbit_drift_flux_expr_2ndform} drops the requirement of
Hamada coordinate. 

The formula being coordinate-choice independent are intuitive because
the flux-force relation and drift orbit flux are physical quantities
that are invariant to how the magnetic surface is parametrized. These
formula could also be preferred in numerical computation because the
construction of straight field line coordinates for a 3D perturbed
plasma is avoided. One can choose the natural poloidal and toroidal
angle $(\theta,\phi)$ as parameters  and the flux label $\psi$ can
be obtained from the magnetic field reconstruction code \cite{Zanca1999,Hanson2009}.

\section{Appendix}

\subsection{Fast calculation by Fourier transformation\label{sub:Fast-calculation-by}}

Evaluation for expressions taking the form 
\begin{equation}
I=\bigl\langle \iint\bigl[\oint h(\vec{x},E,\mu)\frac{dl}{B}\bigr]^{2}dEd\mu\bigr\rangle \label{eq:Fast_fourier_surface}
\end{equation}

can be accelerated with Fourier analysis if certain assumptions are
made. The basic idea is that for a given $(E,\mu)$, the value of
$\oint h\frac{dl}{B}$ is the same for each points on a bounce circle
and may only different by a 'phase' between different field lines.
Thus, if we can factor this $\oint h\frac{dl}{B}$ out and do the
integration with respect to other parameters separately, the calculation
can be simplified a lot. 

To make this improvement, certain assumptions have to be made. We
only consider the trapped particles in this case. The assumptions
are:
\begin{enumerate}
\item The magnetic surface comes from a perturbation to an axisymmetric
$B$ field, and the magnitude of the perturbation is much smaller
to the $|B|$ variation on the original field. 
\item Following the 1st assumption, we further assume that the turning point
for a trapped particle is not changed by the perturbation. 
\end{enumerate}
With such assumptions, if we use the natural poloidal angle $\theta$
as one parameter, and the field line label $\beta$ as another parameter,
then $\oint h\frac{dl}{B}=\int_{\theta_{t1}}^{\theta_{t2}}h\sqrt{g}d\theta$
has the same $\theta_{t1},\theta_{t2}$ on different field lines with
a given $(E,\mu)$. 

One can set the $\theta=0$ line at the most inner board position,
where $|B|$ reaches its maximum. As $\theta$ goes to $2\pi$ after
a poloidal circle, it reaches to the same line. This makes sure that
for any trapped particle trajectory, $\theta$ will always stay within
$[0,2\pi)$ and we don't have to worry about jumps or consistency
issues of $\beta$, as $\theta$ goes beyond $2\pi$. What remains
is to do Fourier transform on $h\sqrt{g}$ w.r.t $\beta$. We only
need the definition of $h\sqrt{g}$ on $(\theta,\beta)\in[0,2\pi)\times[0,2\pi)$
and then make a periodic expansion. 
\begin{eqnarray*}
h\sqrt{g} & = & \sum_{n}(a_{n}(\theta)sin(n\beta)+b_{n}(\theta)cos(n\beta))
\end{eqnarray*}

After changing the order of integration on $d\mu dE$ and the integration
on surface: 
\begin{eqnarray*}
 &  & I=\bigl\langle \iint\bigl[\oint h(\vec{x},E,\mu)\frac{dl}{B}\bigr]^{2}dEd\mu\bigr\rangle \\
 &  & =\iint\bigl(\int\oint\bigl[\oint h(\vec{x},E,\mu)\sqrt{g}d\theta\bigr]^{2}\sqrt{g}d\theta d\beta\bigr)dEd\mu\\
 &  & =\iint\int\sum_{n}\oint\bigl[\tilde{a}_{n}sin(n\beta)+\tilde{b}_{n}cos(n\beta)\bigr]^{2}\sqrt{g}d\theta d\beta dEd\mu
\end{eqnarray*}

where 
\begin{eqnarray*}
\tilde{a}_{n} & = & \int_{\theta_{t1}}^{\theta_{t2}}a_{n}(\theta)d\theta\\
\tilde{b}_{n} & = & \int_{\theta_{t1}}^{\theta_{t2}}b_{n}(\theta)d\theta
\end{eqnarray*}

$\tilde{a}_{n},\tilde{b}_{n}$ are independent of $\theta$ and $\beta$.
If we further assume that the small perturbation to the axisymmetric
equilibrium will result in a $\sqrt{g}=\sqrt{g}(\theta)$ which does
not vary on $\beta$ , then we can integrate on $\beta$ first and
get 
\begin{eqnarray*}
 &  & I=\iint\pi\bigl(\int_{\theta_{t1}}^{\theta_{t2}}\sqrt{g}d\theta\bigr)\sum_{n}(\tilde{a}_{n}^{2}+\tilde{b}_{n}^{2})dEd\mu
\end{eqnarray*}

This expression gets rid of the dimension on $\beta$ and is faster
in computation than \eqref{Fast_fourier_surface}. As for \eqref{orbit_drift_flux_expr},
we have 
\begin{eqnarray*}
 &  & I=\bigl\langle \iint\frac{1}{\mu\nu}|v_{\parallel}|\bigl(\frac{\oint\vec{\nabla}\cdot(\vec{S}|v_{\parallel}|^{3})\frac{dl}{B}}{\oint|v_{\parallel}|\frac{dl}{B}}\bigr)^{2}\bigl(\frac{df_{0}}{d\psi}\bigr)d\mu dE\bigr\rangle \\
 &  & =\iint\frac{1}{\mu\nu}\int\frac{\bigl(\oint\vec{\nabla}\cdot(\vec{S}|v_{\parallel}|^{3})\sqrt{g}d\theta\bigr)^{2}}{\oint|v_{\parallel}|\sqrt{g}d\theta}d\beta\bigl(\frac{df_{0}}{d\psi}\bigr)d\mu dE
\end{eqnarray*}
 We assume that $|v_{\parallel}|\sqrt{g}$ has no $\beta$ dependence
and perform Fourier transform of $\vec{\nabla}\cdot(\vec{S}|v_{\parallel}|^{3})\sqrt{g}$
w.r.t $\beta$ where the dependence of $\beta$ comes from $|B|$
variation. Then 
\[
I=\iint\frac{\pi}{\mu\nu}\frac{\sum_{n}(\tilde{a}_{n}^{2}+\tilde{b}_{n}^{2})}{\oint|v_{\parallel}|\sqrt{g}d\theta}\bigl(\frac{df_{0}}{d\psi}\bigr)d\mu dE
\]
 where $\tilde{a}_{n}$ and $\tilde{b}_{n}$ are the integration on
$\theta$ of the corresponding Fourier coefficients.

If one wants to do the Fourier transform without the involvement of
$v_{\parallel}$(so that FFT will only need to be done once), then
one can use \eqref{orbit_drift_flux_expr_2ndform} and evaluate

\begin{widetext}

\begin{eqnarray*}
 &  & I=\bigl\langle \iint\frac{\mu}{\nu}|v_{\parallel}|\bigl(\frac{\oint\bigl(|v_{\parallel}|\vec{S}\cdot\vec{\nabla}B-\frac{m|v_{\parallel}|^{3}}{3\mu}\vec{\nabla}\cdot\vec{S}\bigr)\frac{dl}{B}}{\oint|v_{\parallel}|\frac{dl}{B}}\bigr)^{2}\bigl(\frac{df_{0}}{d\psi}\bigr)d\mu dE\bigr\rangle \\
 &  & =\iint\frac{\mu}{\nu}\int\frac{\bigl(\oint\bigl(|v_{\parallel}|\vec{S}\cdot\vec{\nabla}B-\frac{m|v_{\parallel}|^{3}}{3\mu}\vec{\nabla}\cdot\vec{S}\bigr)\sqrt{g}d\theta\bigr)^{2}}{\oint|v_{\parallel}|\sqrt{g}d\theta}d\beta\bigl(\frac{df_{0}}{d\psi}\bigr)d\mu dE
\end{eqnarray*}

\end{widetext}

Suppose the Fourier coefficients of $\sqrt{g}\vec{S}\cdot\vec{\nabla}B$
and $\sqrt{g}\vec{\nabla}\cdot\vec{S}$ are: 
\begin{eqnarray*}
\sqrt{g}\vec{S}\cdot\vec{\nabla}B & = & \sum_{n}(a_{n}(\theta)sin(n\beta)+b_{n}(\theta)cos(n\beta))\\
\sqrt{g}\vec{\nabla}\cdot\vec{S} & = & \sum_{n}(c_{n}(\theta)sin(n\beta)+d_{n}(\theta)cos(n\beta))
\end{eqnarray*}

Then we have: 
\[
I=\iint\frac{\pi\mu}{\nu}\frac{\sum_{n}(\tilde{k}_{sin}^{2}+\tilde{k}_{cos}^{2})}{\oint|v_{\parallel}|\sqrt{g}d\theta}\bigl(\frac{df_{0}}{d\psi}\bigr)d\mu dE
\]

where the coefficients are defined as: 
\begin{eqnarray*}
\tilde{k}_{sin} & = & \int_{\theta_{t1}}^{\theta_{t2}}\bigl(|v_{\parallel}|a_{n}(\theta)-\frac{m|v_{\parallel}|^{3}}{3\mu}c_{n}(\theta)\bigr)d\theta\\
\tilde{k}_{cos} & = & \int_{\theta_{t1}}^{\theta_{t2}}\bigl(|v_{\parallel}|b_{n}(\theta)-\frac{m|v_{\parallel}|^{3}}{3\mu}d_{n}(\theta)\bigr)d\theta
\end{eqnarray*}

This gives a boost in computation for the case where the ratio of
perturbation over the axisymmetric equilibrium is much smaller compared
to the aspect ratio $\epsilon$. 
\begin{acknowledgments}
The authors want to thank Andrew J. Cole and Allen H. Boozer for helpful
discussions.

This work was supported by U.S. Department of Energy(DOE) Grant DE
- FG02 - 86ER53222.
\end{acknowledgments}
\bibliographystyle{jphysicsB}

\end{document}